\documentclass{article}
\usepackage{alltt}
\usepackage[all]{xy}
\usepackage{latexsym,amssymb,amsmath,amsfonts, amsthm, xcolor}

\def\vbar{\mathchoice{\vrule height6.3ptdepth-.5ptwidth.8pt\kern-.8pt}
  {\vrule height6.3ptdepth-.5ptwidth.8pt\kern-.8pt}
  {\vrule height4.1ptdepth-.35ptwidth.6pt\kern-.6pt}
  {\vrule height3.1ptdepth-.25ptwidth.5pt\kern-.5pt}}
\def\fudge{\mathchoice{}{}{\mkern.5mu}{\mkern.8mu}}
\def\bbc#1#2{{\rm \mkern#2mu\vbar\mkern-#2mu#1}}
\def\bbb#1{{\rm I\mkern-3.5mu #1}}
\def\bba#1#2{{\rm #1\mkern-#2mu\fudge #1}}
\def\bb#1{{\count4=`#1 \advance\count4by-64 \ifcase\count4\or\bba A{11.5}\or
  \bbb B\or\bbc C{5}\or\bbb D\or\bbb E\or\bbb F \or\bbc G{5}\or\bbb H\or
  \bbb I\or\bbc J{3}\or\bbb K\or\bbb L \or\bbb M\or\bbb N\or\bbc O{5} \or
  \bbb P\or\bbc Q{5}\or\bbb R\or\bbc S{4.2}\or\bba T{10.5}\or\bbc U{5}\or
  \bba V{12}\or\bba W{16.5}\or\bba X{11}\or\bba Y{11.7}\or\bba Z{7.5}\fi}}

\newcommand{\RR}{\mbox{${\rm \:  R\!\!\!\! I
\;\;}$}}

\newcommand{\vs}{\vspace{0.25cm}}

\newtheorem{theorem}{Theorem}
\newtheorem{itlemma}{Lemma}[section]
\newtheorem{itproposition}[itlemma]{Proposition}
\newtheorem{itcorollary}[itlemma]{Corollary}
\newtheorem{itremark}[itlemma]{Remark}
\newtheorem{itremarks}[itlemma]{Remarks}
\newtheorem{itdefinition}[itlemma]{Definition}
\newtheorem{itexample}[itlemma]{Example}

\newenvironment{lemma}{\begin{itlemma}\rm}{\end{itlemma}} 
\newenvironment{remark}{\begin{itremark}\rm}{\end{itremark}} 
\newenvironment{remarks}{\begin{itremarks} \rm}{\end{itremarks}}
\newenvironment{corollary}{\begin{itcorollary}\rm}{\end{itcorollary}}
\newenvironment{proposition}{\begin{itproposition}\rm}{\end{itproposition}}
\newenvironment{definition}{\begin{itdefinition}\rm}{\end{itdefinition}}
\newenvironment{example}{\begin{itexample}\rm}{\end{itexample}}
\newenvironment{fact}{\noindent {{\bf Fact}}:\ \ }{\hfill \medskip}
\newenvironment{claim}{\noindent {\em Claim}. \ \ }{\hfill \medskip}
\newcommand{\be}[1]{\begin{equation}\label{#1}}
\newcommand{\ee}{\end{equation}}
\newcommand{\bl}[1]{\begin{lemma}\label{#1}}
\newcommand{\br}[1]{\begin{remark}\label{#1}}
\newcommand{\brs}[1]{\begin{remarks}\label{#1}}
\newcommand{\bt}[1]{\begin{theorem}\label{#1}}
\newcommand{\bd}[1]{\begin{definition}\label{#1}}
\newcommand{\bp}[1]{\begin{proposition}\label{#1}}
\newcommand{\bc}[1]{\begin{corollary}\label{#1}}
\newcommand{\bfact}[1]{\begin{fact}\label{#1}}
\newcommand{\bex}[1]{\begin{example}\label{#1}}
\newcommand{\ec}{\end{corollary}}
\newcommand{\efact}{\end{fact}}
\newcommand{\eex}{\end{example}}
\newcommand{\el}{\end{lemma}}
\newcommand{\er}{\end{remark}}
\newcommand{\ers}{\end{remarks}}
\newcommand{\et}{\end{theorem}}
\newcommand{\ed}{\end{definition}}
\newcommand{\ep}{\end{proposition}}
\newcommand{\epr}{\end{proof}}
\newcommand{\bpr}{\begin{proof}}
\newcommand{\bcl}{\begin{claim}}
\newcommand{\ecl}{\end{claim}}

\newcommand{\bi}{\begin{itemize}}
\newcommand{\ei}{\end{itemize}}
\newcommand{\ben}{\begin{enumerate}}
\newcommand{\een}{\end{enumerate}}

\usepackage{graphicx}

\title{\bf \Large{Controllability of Symmetric Spin Networks}}

\vs

\vs

\author{Francesca Albertini\thanks{Dipartimento di Tecnica e Gestione 
dei Sistemi Industriali,  Universit\`a di Padova, albertin@math.unipd.it}  \, \, \,  and \, Domenico D'Alessandro\thanks{Department of Mathematics, Iowa State University, Ames, Iowa, U.S.A., e-mail:daless@iastate.edu}}

\begin{document}

\maketitle

\begin{abstract} We consider a network of $n$ spin $\frac{1}{2}$ systems which are pairwise interacting  via Ising interaction and are controlled  by the same electro-magnetic control field. Such a system presents symmetries since the Hamiltonian is unchanged if we permute two spins. This prevents full (operator) controllability \cite{Mikobook} in that not every  unitary evolution can be obtained. We prove however that controllability is verified if we restrict ourselves to unitary evolutions
 which preserve the above permutation invariance. For low dimensional cases, $n=2$ and $n=3$, we provide an analysis of the Lie group of available evolutions and give explicit control laws to transfer between any two permutation invariant states. This class of states includes highly entangled states such as $GHZ$ states \cite{GHZ1} and $W$ states \cite{Wref1}, which are of 
 interest in quantum information.  
\end{abstract}

\vs 
\vs 
\vs


\vs 
\vs 
\vs 

\section{Introduction} 

The {\it controllability} of a control system describes the  set of states which can be reached for that system by opportunely changing the external controls. For finite dimensional quantum 
systems, controllability is usually assessed by calculating the Lie algebra generated by the Hamiltonians of the system \cite{Mikobook}. Such a Lie algebra is called the {\it dynamical Lie algebra}. If the dynamical Lie algebra of a system of dimension $d$ is the full Lie algebra $u(d)$ ($su(d)$) of $d\times d$ skew-Hermitian matrices (with zero trace) then the set of available evolutions is the full Lie group of $d \times d$ unitary matrices $U(d)$ (with determinant equal to $1$, $SU(d)$) and the system is said to be {\it operator controllable}.\footnote{The term `completely controllable' is used to denote this situation \cite{SophieSchirmer}.}  More in general the set of the available evolutions is dense in the Lie group associated with the dynamical Lie algebra and it is equal to such a Lie group in the case where such a Lie group is compact.  Although controllability is the generic situation \cite{Lloyd}, in reality, symmetries present in the system's dynamics restrict the type of available evolutions. In this paper, we analyze one of these situations for a system of interest in the implementation of quantum information processing and the generation of entangled states. 

In particular, we consider a network of $n \geq 2$ spin controlled in parallel 
by an electromagnetic field. Such a system was also considered in \cite{Xinhua} and allows one to perform quantum information processing and generate entangled states without the need to 
address the spins individually.  A common control field is used to control 
all spins {\it simultaneously}. The   Hamiltonian (cf. (\ref{basicHam}) below) is symmetric in the sense that is invariant under permutation. As a consequence, starting from the ground state where all the spin are in state $0\rangle$ the possible states of the spin network will also be permutation invariant. Examples of such states are the GHZ states introduced in \cite{GHZ1}
\be{GHZ2}
|GHZ \rangle:=\frac{1}{\sqrt{2}}(|000\cdots 0 \rangle +|111\cdots 1 \rangle),  
\ee  
and the W states 
\be{Wstates}
|W\rangle := \frac{1}{\sqrt{n}}( |100\cdots 0\rangle+|010\cdots 0\rangle+ 
|001\cdots 0\rangle+\cdots + |000\cdots 1\rangle), 
\ee
considered in \cite{Wref1}. However,  we do not restrict ourselves to transfer to or from these states as in \cite{Xinhua} but consider the control problem on the full subspace of permutation invariant states. A basis for such a subspace is given by the $n+1$ orthonormal states, for $m=0,1,...,n$, 
\be{statesphi}
|\phi_m\rangle =\frac{1}{\sqrt{n \choose m}}\sum_k |k\rangle. 
\ee 
In formula (\ref{statesphi}), for each $m$ the sum runs for all the elements in the computational basis which have $m$ $1$'s and $n-m$ $0$'s. For example, $ |GHZ \rangle $ 
in (\ref{GHZ2}) is $\frac{1}{\sqrt 2} (|\phi_0 \rangle + |\phi_n \rangle)$, while $|W \rangle$ in (\ref{Wstates}) is $|\phi_1 \rangle$. 

\vs 

We {\bf present  the following results:}

\begin{enumerate}

\item We prove the controllability  of the system in the sense that 
the dynamical Lie algebra ${\cal L}$ (see, e.g., \cite{Mikobook})  is the full Lie subalgebra of $u(2^n)$ consisting of matrices which have zero trace and are   permutation invariant. The corresponding Lie group is compact and therefore  the set of possible available evolutions is equal to the corresponding Lie group $e^{\cal L}$ (Theorem \ref{Theo1B}).

\vs  

{\bf For the cases $n=2$ and $n=3$:}

\vs

\item  We explicitly describe the dynamical Lie algebra ${\cal L}$ 
which is, after a change of coordinates, spanned by matrices in a 
 direct sum of $u(1)$ with $u(3)$ and with zero trace for the case $n=2$. It is spanned by matrices in  a direct sum of two copies of (a Lie algebra isomorphic to) $u(2)$ and  $u(4)$ and with trace equal to zero for the case $n=3$. From this, a parametrization of the Lie group of the available evolutions is obtained and controllability between any two permutation invariant pure states is proved. Results are summarized in Theorem \ref{TheosumN2} and \ref{TheosumN3} for the cases $n=2$ and $n=3$, respectively.

\item We use the above parametrization of 
the Lie group of available evolutions and Lie group decomposition techniques to 
provide explicit control algorithms to transfer between 
any two  permutation invariant states (Section \ref{Algo1}).

\end{enumerate}

\subsection{Definitions  and elementary properties}

The Pauli matrices $\sigma_{x,y,z}$ are defined as 
\be{PauliMat}
\sigma_x:=\begin{pmatrix} 0 & 1 \cr 1 & 0  \end{pmatrix}, \qquad \sigma_y:=
 \begin{pmatrix} 0 & i \cr -i & 0  \end{pmatrix}, \qquad
 \sigma_z:= \begin{pmatrix}  1 & 0 \cr 0 & -1\end{pmatrix}.
\ee

Let $\bf{1}$ be the two dimensional identy matrix, the Pauli matrices  satisfy 
\be{relations}
\begin{array}{c}
\sigma_x\sigma_x=\sigma_y\sigma_y=\sigma_z\sigma_z={\bf 1}, \\
\begin{array}{ccc}
\sigma_x\sigma_y=-i \sigma_z,&  \sigma_y\sigma_z=-i \sigma_x, & \sigma_z\sigma_x=-i \sigma_y \\
\sigma_y\sigma_x=i \sigma_z,& \sigma_z\sigma_y=i \sigma_x, &  \sigma_x\sigma_z=i \sigma_y \\
\end{array} \\
\end{array}
\ee

{The quantum system we study in this paper is a symmetric Ising spin chain under the control of a common electromagnetic field.  The corresponding, time varying controlled Hamiltonian, for  $n \geq 2$ spin $\frac{1}{2}$ particles is defined as:
\be{basicHam}
H(t)=H_{zz}+H_x u_{x}(t) +H_yu_{y}(t), 
\ee
where 
\be{basicHamdetails}
\begin{array}{l}
H_{zz}= \sum_{1\leq k<m\leq n} 
{\bf{1}}\otimes\cdots{\bf{1}}\otimes\underbrace{\sigma_z}_{k^{th} }\otimes{\bf{1}}\otimes\cdots{\bf{1}}
\otimes\underbrace{\sigma_z}_{m^{th} }\otimes{\bf{1}}\otimes\cdots{\bf{1}}, \\
 H_{x}=\sum_{k=1}^{n} {\bf{1}}\otimes\cdots{\bf{1}}\otimes\underbrace{\sigma_x}_{k^{th} }\otimes{\bf{1}}\otimes\cdots{\bf{1}}, \\
 H_{y}=\sum_{k=1}^{n} {\bf{1}}\otimes\cdots{\bf{1}}\otimes\underbrace{\sigma_y}_{k^{th} }\otimes{\bf{1}}\otimes\cdots{\bf{1}}. \end{array}
\ee 

Here $u_x$ and $u_y$ represent the $x$ and $y$ components of a control electromagnetic field. $H_{zz}$ is the interaction Hamiltonian which is the sum of $n \choose 2$ Ising interactions between all pairs of spins. The Hamiltonians $H_x$, 
$H_y$ 
model the interaction of the spins with the external control field.    
 
It is clear that the given Hamiltonian is invariant under a permutation of the spins, we next define the Lie subalgebra of $u(2^n)$ of matrices which are permutation invariant.
The $4 \times 4$  matrix 
\be{Pimat}
\Pi:=\begin{pmatrix}1 & 0 & 0 & 0 \cr 
0 & 0 & 1 & 0 \cr 
0& 1 & 0 & 0 \cr 
0 & 0 & 0 & 1 \end{pmatrix}, 
\ee 
is such that for any two, $2 \times 2$  matrices,  $\sigma_1$ and $\sigma_2$ 
\be{PS1S2}
\Pi \, \sigma_1 \otimes \sigma_2\,  \Pi=\sigma_2 \otimes \sigma_1.  
\ee
The $n-1$ matrices $\Pi_{j,j+1}$, $j=1,...,n-1$,  defined as 
\be{uij}
\Pi_{j,j+1}:={\bf 1}^{\otimes j-1} \otimes \Pi \otimes {\bf 1}^{\otimes n-j-1}, 
\ee
generate the whole group of permutations in the sense that every 
permutation of positions of the $n$ factors in a tensor product can be obtained by multiplications of such matrices. Therefore the Lie subalgebra of $u(2^n)$ of matrices which are permutation invariant can be described as 
\be{LPI}
{\cal L}^{PI}:=\{A \in u(2^n) \, | \, \Pi_{j,j+1} A \Pi_{j,j+1}, \, j=1,\ldots n-1 \}. 
\ee 
Since $\Pi^2={\bf 1} \otimes {\bf 1}$, we have that 
$\Pi_{j,j+1}^2={\bf 1}^{\otimes n}$, for every $j=1,1,...,n-1$. From this it follows that  if  general $2 ^n \times 2^n$ skew-Hermitian matrices $A$ and $B$ are such $\Pi_{j,j+1}A\Pi_{j,j+1}=A$ and $\Pi_{j,j+1}B\Pi_{j,j+1}=B$, $AB$ is also such that $\Pi_{j,j+1}AB\Pi_{j,j+1}=\Pi_{j,j+1}A\Pi_{j,j+1}^2B\Pi_{j,j+1}=AB$.  This confirms that ${\cal L}^{PI}$ is a Lie subalgebra of $u(2^n)$, since it is closed under commutation.

\section{Controllability}

Applying general results on the controllability of systems on Lie groups \cite{JS} and quantum systems (see, e.g., \cite{Mikobook}, \cite{HT}, \cite{Salapaka})  the study of the controllability of system (\ref{basicHam}) will be carried out  
by evaluating the dynamical Lie algebra ${\cal L}$ in $u(2^n)$ generated 
by the matrices $\{iH_{zz},iH_x,iH_y\}$ in (\ref{basicHam}). It is known that 
the set of reachable evolutions is dense in the Lie group, $e^{\cal L}$, associated with the dynamical Lie algebra ${\cal L}$ and coincides with such a Lie group if this Lie group is compact. The following result characterizes the dynamical Lie algebra for system (\ref{basicHam}). It shows that, except for the fact that the matrices corresponding to $H_{zz}$, $H_x$ and $H_y$ in (\ref{basicHamdetails}) have zero trace, the dynamical Lie algebra is the full Lie algebra of permutation invariant skew-Hermitian matrices in $u(2^n)$.  

\bt{Theo1} (Dynamical Lie Algebra) 
The dynamical Lie algebra ${\cal L}$ associated 
with system (\ref{basicHam}) coincides with the 
space of all permutation invariant matrices in
 $su(2^n)$, i.e., with ${\cal L}^{PI}$ in 
(\ref{LPI}), 
\be{stateTheo}
{\cal L}={\cal L}^{PI} \cap su(2^n).
\ee
\et 

\vs 

The set of possible evolutions for the system (\ref{basicHam}) (\ref{basicHamdetails}) is described in the following theorem. 
 
\bt{Theo1B}  (Controllability) 
The set of possible evolutions for the system characterized by the Hamiltonian (\ref{basicHam}) is the compact Lie group corresponding to the Lie algebra ${\cal L}^{PI} \cap  su(2^n)$.  
\et  
\bpr 
The claim follows from the standard results on controllability of right invariant systems on Lie groups 
\cite{JS} and quantum mechanical systems \cite{Mikobook}, \cite{HT}, \cite{Salapaka} and the observation that $e^{{\cal L}^{PI}}$ is compact (which is proved in  Remark \ref{compactness} below) and $SU(2^n)$ is 
also compact, so that 
$$
e^{{\cal L}^{PI} \cap su(2^n)}=e^{{\cal L}^{PI}} \cap SU(2^n), 
$$
is also compact. From known results for right invariant systems 
on Lie groups \cite{JS} (because of compactness) the set of 
available evolutions is exactly equal to such a Lie group. 
\epr

We  now calculate the dimension  of the Lie algebra ${\mathcal{L}}$ in (\ref{stateTheo}). To do this we 
first introduce some notations.
 Let $\sigma_0:={\bf 1}$. 
For an $n-$ple $\underline{l}:=(l_1,l_2,...,l_n)$ of elements in the set $\{0,x,y,z\}$, 
we denote by 
\be{sigund}
\sigma_{\underline{l}}:=\sigma_{l_1}\otimes\cdots\otimes \sigma_{l_n}.
\ee

Consider a matrix 
$X$ of the type $X=i\sum_{\underline{l}}\alpha_{\underline{l}}\sigma_{\underline{l}}\in su(2^n)$, where the sum is taken among the $4^n$ possible $n-$ples in $\{0,x,y,z\}$, with coefficients $\alpha_{\underline l}$.  For any permutation $\pi$ we let
$X^{\pi}=i\sum_{\underline{l}}\alpha_{\underline{l}}\sigma_{\pi(\underline{l})}$.
Thus  $X$ is  permutation invariant, i.e. $X\in {\mathcal L}^{PI}$ (see equation (\ref{LPI})), if and only if  $X=X^{\pi}$ for all permutations $\pi\in S_n$.


For a triple $(k_x,k_y,k_z)$  indicating the numbers of $\{x,y,z\}$ symbols, we denote by $\Phi (k_x,k_y,k_z)$ the set of $n$-ples with $k_x$, $x's$, $k_y$, $y's$, and $k_z$, $z's$. We let, with definition (\ref{sigund}),  
\[
X^n_{(k_x,k_y,k_z)}=i\sum_{ \underline{l}\in\Phi (k_x,k_y,k_z)}  \sigma_{\underline l}.
\]
Then $X^n_{(k_x,k_y,k_z)}$ is a permutation invariant matrix, and any permutation invariant matrix can be written as a linear combination
of matrices of the type $X^n_{(k_x,k_y,k_z)}$. 

With these notations, in particular we have: 
\be{add1}
iH_{zz}=X^n_{(0,0,2)}, \ \  \ iH_{x}=X^n_{(1,0,0)}, \ \  \  iH_{y}=X^n_{(0,1,0)}.
\ee

We first  calculate the dimension  of the Lie algebra ${\cal L}^{PI}$ in (\ref{LPI}). Elements forming a  basis for the Lie 
algebra ${\cal L}^{PI}$ in (\ref{LPI}) are in one to one correspondence with triple $(k_x,k_y,k_z)$ where $k_{x,y,z}$ denote the number of matrices $\sigma_{x,y,z}$ present on the given element. Therefore, the dimension of ${\cal L}^{PI}$ is given by all the ways to choose the triples $(k_x,k_y,k_z)$, 
with $0 \leq k_x+k_y+k_z \leq n$. 

Now, given $0 \leq k \leq n$, we have $\frac{(k+1)(k+2)}{2}$ ways of choosing  $(k_x,k_y,k_z)$, with $k=k_x+k_y+k_z$. In fact, $k_x$ can be chosen in $k+1$ ways, then $k_y$ can be chosen in $k+1-k_x$ ways, while $k_z=k-(k_x+k_y)$ is now fixed.  Thus by varying $k_x$, from $0$ to $k$,  we have that  the total ways are:
\[
\sum_{k_x=0}^k (k+1-k_x)= \sum_{l=1}^{k+1} l= \frac{(k+1)(k+2)}{2}.
\]
The dimension of the Lie algebra ${\cal L}^{PI}$   in (\ref{LPI})  is obtained by summing the above numbers as $k=0,1,...,n$. 
We have 
\be{dimensione}
{\text{ dim }{\cal L}^{PI}} \,=\,  \sum_{k=0}^n \frac{(k+1)(k+2)}{2}= \frac{1}{2} \sum_{k=0}^n {(k+1)(k+2)}=  \frac{1}{2}\frac{(n+3)(n+2)(n+1)}{3}:={{n+3} \choose n}.
\ee
where the second equality is  proved by induction on $n$. 

Therefore,  the 
 dimension  of the Lie algebra ${\mathcal{L}}$ in (\ref{stateTheo}) is ${{n+3} \choose n  }-1$.

\subsection{Proof of Theorem \ref{Theo1}}

The dynamical  Lie algebra ${\cal L}$ is  generated by $iH_{zz},\,  iH_{x},$ and $iH_{y}$ and since $iH_{zz},\,  iH_{x},$ and $iH_{y}$ belong to ${\cal L}^{PI} \cap su(2^n)$, ${\cal L} \subseteq  {\cal L}^{PI} \cap su(2^n)$.  
To prove Theorem \ref{Theo1} we need to establish also the converse inclusion, i.e.  ${\mathcal L}^{PI} \cap su(2^n) \subseteq {\mathcal{L}}$.
To get this inclusion, we 
 will prove that,  
\be{formula}
\forall \, (k_x,k_y,k_z) \text{ such that } 1\leq k_k+k_y+k_z\leq n,   \, X^n_{(k_x,k_y,k_z)}\in {\mathcal{L}}. 
\ee
 We will prove equation (\ref{formula}) by induction on $\bar{k}=k_x+k_y+k_z$ ($1\leq \bar{k}\leq n$). We will derive equation (\ref{formula}) for $\bar{k}=1,\, 2$ first and then will prove the inductive step. 
 
\begin{itemize} 
\item $\bar k=1$. 

For $\bar k=1$ there are only three possible triples $(k_x,k_y,k_z)$, with $k_x+k_y+k_z=\bar k$. 
$X^n_{(1,0,0)}$ and $X^n_{(0,1,0)}$ are already in $\cal{L}$ because of (\ref{add1}). Moreover a direct calculation gives 
$$
[X^n_{(1,0,0)},X^n_{(0,1,0)}]=2X^n_{(0,0,1)}, 
$$
since 
\[
[X^n_{(1,0,0)},X^n_{(0,1,0)}]=-\sum_{i,j=1}^n [{\bf{1}}\otimes\cdots{\bf{1}}\otimes\underbrace{\sigma_x}_{i^{th} }\otimes{\bf{1}}\otimes\cdots{\bf{1}},\, 
{\bf{1}}\otimes\cdots{\bf{1}}\otimes\underbrace{\sigma_y}_{j^{th} }\otimes{\bf{1}}\otimes\cdots{\bf{1}}]=
\]
\[
=2i \sum_{i}^n {\bf{1}}\otimes\cdots{\bf{1}}\otimes\underbrace{\sigma_z}_{i^{th} }\otimes{\bf{1}}\otimes\cdots{\bf{1}},
\]
since we have that, if $i\neq j$ then the two matrices commute, and if $i=j$ then  $[\sigma_x,\sigma_y]=-2i\sigma_z$.

\item $\bar k=2$. 

For $\bar k=2$ there are only $6$ possible triples $(k_x, k_y, k_z)$ 
with $k_x+k_y+k_z=\bar k$. $X^n_{(0,0,2)}$ is already 
in ${\cal {L}}$ because of (\ref{add1}). Moreover, we calculate 
\be{l2}
\begin{array}{ccc}
[ X^n_{(1,0,0)},X^n_{(0,0,2)} ]  &  = & -2X^n_{(0,1,1)}\\
{[ X^n_{(0,1,0)},X^n_{(0,0,2)} ]} & =& 2X^n_{(1,0,1)}
\end{array}
\ee
Let us give details on the first equation.  The second one is similar. 
First we notice that:
\[
[{\bf{1}}\otimes\cdots{\bf{1}}\otimes\underbrace{\sigma_x}_{i^{th} }\otimes{\bf{1}}\otimes\cdots{\bf{1}},\, {\bf{1}}\otimes\cdots{\bf{1}}\otimes\underbrace{\sigma_z}_{j^{th} }\otimes\cdots\otimes\underbrace{\sigma_z}_{l^{th} }\otimes\cdots {\bf{1}} \otimes \cdots{\bf{1}}]=
\]
\be{l21}
=\left\{\begin{array}{cc}
0   & \text{ if $i\neq j$ and $i\neq l$}\\
2i \ {\bf{1}}\otimes\cdots{\bf{1}}\otimes\underbrace{\sigma_y}_{j^{th} }\otimes\cdots\otimes\underbrace{\sigma_z}_{l^{th} }\otimes\cdots {\bf{1}} \otimes \cdots{\bf{1}}
&  \text{ if $i=j$} \\
2i \ {\bf{1}}\otimes\cdots{\bf{1}}\otimes\underbrace{\sigma_z}_{j^{th} }\otimes\cdots\otimes\underbrace{\sigma_y}_{l^{th} }\otimes\cdots {\bf{1}} \otimes \cdots{\bf{1}}
&  \text{ if $i=l$} 
\end{array}\right.
\ee
Thus:
\[
[ X^n_{(1,0,0)},X^n_{(0,0,2)} ] =-\sum_{\begin{array}{l} i,j,l=1\\ j<l \end{array}}^n [{\bf{1}}\otimes\cdots{\bf{1}}\otimes\underbrace{\sigma_x}_{i^{th} }\otimes{\bf{1}}\otimes\cdots{\bf{1}},\, {\bf{1}}\otimes\cdots{\bf{1}}\otimes\underbrace{\sigma_z}_{j^{th} }\otimes\cdots\otimes\underbrace{\sigma_z}_{l^{th} }\otimes\cdots {\bf{1}} \otimes \cdots{\bf{1}}]=
\]
\[=
-\sum_{\begin{array}{l} j,l=1\\ j<l \end{array}}^n 2i \left( \ {\bf{1}}\otimes\cdots{\bf{1}}\otimes\underbrace{\sigma_y}_{j^{th} }\otimes\cdots\otimes\underbrace{\sigma_z}_{l^{th} }\otimes\cdots {\bf{1}} \otimes \cdots{\bf{1}} + 
 {\bf{1}}\otimes\cdots{\bf{1}}\otimes\underbrace{\sigma_z}_{j^{th} }\otimes\cdots\otimes\underbrace{\sigma_y}_{l^{th} }\otimes\cdots {\bf{1}} \otimes \cdots{\bf{1}} \right)=
\]
\[ =
-2 X^n_{(0,1,1)}.
\]

This shows that $X^n_{(0,1,1)}$ and $X^n_{(1,0,1)}$ are in ${\cal L}$. Moreover we have with similar calculations (cf. Appendix A)
\be{forApp}
\begin{array}{ccc}
[ X^n_{(0,1,1)},X^n_{(1,0,0)} ]  &  = &  4 X^n_{(0,2,0)} - 4 X^n_{(0,0,2)}\\
{[ X^n_{(1,0,1)},X^n_{(0,1,0)} ]}  &  = & -4 X^n_{(2,0,0)} +4 X^n_{(0,0,2)}, 
\end{array}
\ee
and therefore $X^n_{(0,2,0)}$, and $X^n_{(2,0,0)}$ are  also in ${\cal{L}}$. 
Finally, using a similar argument as the one used to prove  (\ref{l2}), we have, 
\be{l3-bis}
{[ X^n_{(0,0,1)},X^n_{(2,0,0)} ]}  = 2\, X^n_{(1,1,0)}. 
\ee
Therefore all the basis matrices corresponding to $\bar k=2$ are in ${\cal L}$. 

\item By exchanging the roles of $x$, $y$ and $z$, we can see the following: 

{\bf Fact:}
Assume that $X^n_{(k_x,k_y,k_z)}\in {\mathcal{L}}$  for  
all triples $(k_x,k_y,k_z)$, with $k_x+k_y+k_z\leq \hat{k}$ and that 
$X^n_{(\tilde{k}_x,\tilde{k}_y,\tilde{k}_z)}$, with $\tilde{k}_x+\tilde{k}_x+\tilde{k}_x>\hat{k}$ is obtained as a Lie bracket of elements $X^n_{(k_x,k_y,k_z)}$ with $k_x+k_y+k_z=\hat{k}$, and therefore is in ${\cal L}$. Then, every $X^n_{(\hat{k}_x,\hat{k}_y,\hat{k}_z)}$ is also in ${\cal L}$, where $(\hat{k}_x,\hat{k}_y,\hat{k}_z)$ is any 
 permutation of $(\tilde{k}_x,\tilde{k}_y,\tilde{k}_z)$.

\item Inductive step: Let $2\leq\bar{k}-1<n$ and assume that  all possible  $X^n_{(k_x,k_y,k_z)}\in {\mathcal{L}}$ for $1\leq k_x+k_y+k_z\leq \bar{k}-1$.  
Then also all $X^n_{(k_x,k_y,k_z)}\in {\mathcal{L}}$ with 
$k_x+k_y+k_z=\bar{k}$.

\bpr
By the symmetry property of the above {\bf Fact}, it is enough to prove that, using Lie bracket of elements 
$X^n_{(k_x,k_y,k_z)}\in {\mathcal{L}}$ with $1 \leq k_x+k_y+k_z\leq \bar{k}-1$, we can obtain all $X^n_{(k_x,k_y,k_z)}$ for $k_x+k_y+k_z=\bar{k}$ {\it with the restriction that $k_z\leq k_y\leq k_x$.} Such a restriction does not imply a loss of generality. 

We set $k_x=\bar{k}-j$ and will prove this fact by induction on $j$. The possible range of values for $j$ is $0\leq j\leq [\frac{2\bar{k}}{3}]$.  In fact, if $j= [\frac{2\bar{k}}{3}]+1$, then in particular $j>\frac{2\bar{k}}{3}$, thus $k_x=\bar{k}-j<\bar{k} -\frac{2\bar{k}}{3}= \frac{\bar{k}}{3}$, so also
$k_x+k_y+k_z\leq 3k_x<\bar{k}$. On the other hand if $j= [\frac{2\bar{k}}{3}]$, then $k_x=\bar{k}-j\geq \frac{\bar{k}}{3}$, thus $3k_x\geq \bar{k}$, so there exists a triple 
$(\bar{k}-j,k_y,k_z)$, with $\bar{k}-j+k_y+k_z=\b ar{k}$ and $k_z\leq k_y\leq \bar{k}-j$

{\em{Base step}}: $j=0$ $\Rightarrow$ $k_x=\bar{k}$.

To get this base step, $j=0$, we will prove also the cases $j=1$ and $j=2$.
It holds that:
\be{e0}
[X^n_{(\bar{k}-1,0,0)},X^n_{(1,1,0)}]= 
2X^n_{(\bar{k}-1,0,1)}+2 X^n_{(\bar{k}-2,0,1)}.
\ee
To see this, assume we have an $n-$ple   $\underline{i}$ which has $\bar{k}-1$ elements equal to $x$   and all the other equal to $0$ and an $n-$ple  $\underline{j}$ with one element
 $j_x=x$ one  $j_y=y$ and all the other equal to $0$. Denote by $A$ the set of indexes such that $i_l=x$. 
We have:
\be{e00}
[\sigma_{\underline{i}}, \sigma_{\underline{j}}] = \left\{\begin{array}{cl}
0 &   { \text{if $j_y\not \in A$}} \\
-2i \sigma_{l_1}\otimes\cdots\otimes\sigma_{l_n}  & {\text{if $j_x \in A$ and $j_y  \in A$}}   \\
-2i \sigma_{s_1}\otimes\cdots\otimes\sigma_{s_n}  &   {\text{if $j_x \not \in A$ and $j_y \in A$}}  
\end{array} \right.
\ee
where the $n-$ple $\underline{l}=(l_1,\ldots, l_n)$ has $\bar{k}-2$ indexes equal to $x$ and one equal to $z$, while the $n-$ple  $\underline{s}=(s_1,\ldots, s_n)$ has $\bar{k}-1$ indexes equal to $x$ and one equal to $z$.
Since $[X^n_{(\bar{k}-1,0,0)},X^n_{(1,1,0)}]$ is a permutation invariant matrix, and $X^n_{(\bar{k}-1,0,0)}$ is a sum of all elements of the type $\sigma_{\underline{i}}$, while 
$X^n_{(1,1,0)}$ is a sum of all elements of the type $\sigma_{\underline{j}}$, from (\ref{e00}), 
equation (\ref{e0}) follows.

From equation (\ref{e0}) since $X^n_{(\bar{k}-2,0,1)}$ is in ${\mathcal{L}}$, we have 
that  $X^n_{(\bar{k}-1,0,1)}\in {\mathcal{L}}$.
By the symmetry property of the above  {\bf Fact} we also have  $X^n_{(\bar{k}-1,1,0)}\in {\mathcal{L}}$. The next two equations can be proved by direct calculation:
\be{e1}
[X^n_{(\bar{k}-1,1,0)},X^n_{(0,0,1)}]= 
-2X^n_{(\bar{k}-2,2,0)}+2X^n_{(\bar{k},0,0)}.
\ee
\be{e2}
[X^n_{(\bar{k}-1,0,1)},X^n_{(0,1,0)}]= 
2X^n_{(\bar{k}-2,0,2)}-2X^n_{(\bar{k},0,0)}.
\ee
We now compute $[X^n_{(\bar{k}-2,1,0)},X^n_{(1,0,1)}]$ with an argument 
similar to the one used to derive equation (\ref{e0}). 
Assume we have an $n-$ple   $\underline{i}$ which has $\bar{k}-2$ elements equal to $x$, one element $i_y=y$, and all the other equal to $0$,  and an $n-$ple  $\underline{j}$ with one element $j_x=x$ one  $j_z=z$ and all the other equal to $0$. Denote by $A$ the set of indexes such that $i_l=x$. 
We have:
\be{e30}
[\sigma_{\underline{i}}, \sigma_{\underline{j}}] = \left\{\begin{array}{cl}
0 &   {\text{if }}\begin{array}{l}  j_x\not \in A\cup\{i_y\}  \text{ and  } j_z\not \in A\cup\{i_y\}  \text{ or } \\
                                         j_x \in A \text{ and  } j_z\not \in A\cup\{i_y\}    \text{ or } \\
                                          j_x=i_y \text{ and  } j_z \in A \end{array} \\
-2i \sigma_{l_1}\otimes\cdots\otimes\sigma_{l_n}  & {\text{if $j_x \not\in A\cup\{i_y\}$ and $j_z=i_y$}}   \\
2i \sigma_{s_1}\otimes\cdots\otimes\sigma_{s_n}  &   {\text{if $j_x \not\in A\cup\{i_y\}$ and $j_z \in A$}} \\
 2i \sigma_{m_1}\otimes\cdots\otimes\sigma_{m_n}  & {\text{if $j_x \in A$ and $j_z \in A$}}   \\
 -2i \sigma_{q_1}\otimes\cdots\otimes\sigma_{q_n}  &   {\text{if $j_x \in A$ and $j_z =i_y$}}  \\
  2i \sigma_{r_1}\otimes\cdots\otimes\sigma_{r_n}  &   {\text{if $j_x =i_y$ and $ j_z\not \in A\cup\{i_y\}$}}.  \\
\end{array} \right.
\ee
Here the $n-$ple  $\underline{l}=(l_1,\ldots, l_n)$ has $\bar{k}$ indexes equal to $x$, the $n$-ple $\underline{s}=(s_1,\ldots, s_n)$ has $\bar{k}-2$ indexes equal to $x$ and two equal to $y$,  the $n-$ple  $\underline{m}=(m_1,\ldots, m_n)$ has $\bar{k}-3$ indexes equal to $x$ and two equal to $y$,  the $n$-ple  $\underline{q}=(q_1,\ldots, q_n)$ has $\bar{k}-2$ indexes equal to $x$, and  the $n$-ple $\underline{r}=(r_1,\ldots, r_n)$ has $\bar{k}-2$ indexes equal to $x$ and two equal to $z$.
Since $X^n_{(\bar{k}-2,1,0)}$ is a sum of all elements of the type $\sigma_{\underline{i}}$, while 
$X^n_{(1,0,1)}$ is a sum of all elements of the type $\sigma_{\underline{j}}$, from (\ref{e30}), we have: 
\be{e3}
[X^n_{(\bar{k}-2,1,0)},X^n_{(1,0,1)}]=
-2 X^n_{(\bar{k}-3,2,0)}+2 X^n_{(\bar{k}-2,0,0)}-2 X^n_{(\bar{k}-2,2,0)}-2X^n_{(\bar{k}-2,0,2)}+2X^n_{(\bar{k},0,0)}.
\ee
Since $X^n_{(\bar{k}-3,2,0)}$ and $X^n_{(\bar{k}-2,0,0)}$ are in ${\mathcal{L}}$, 
putting together equations (\ref{e1}), (\ref{e2}), and (\ref{e3}), we get that
\[
X^n_{(\bar{k},0,0)}, \,  X^n_{(\bar{k}-2,2,0)}, \, X^n_{(\bar{k}-2,0,2)} \in {\mathcal{L}}^n.
\]
So, in particular,  we have proven that ${\mathcal{L}}$ contains all
$X^n_{(\bar{k},0,0)}$, which is the base step, $j=0$.

{\em{Induction step}}:
Assume we have in ${\cal L}$ all $X^n_{(k_x,k_y,k_z)}$ with 
$k_x+k_y+k_z=\bar{k}$ and $k_x=\bar{k}-(j-1)$, we want to show that we also have all $X^n_{(k_x,k_y,k_z)}$ with 
$k_x=\bar{k}-j$.

Fix a triple $k_x+k_y+k_z=\bar{k}$, with $k_x=\bar{k}-j$. Certainly $k_y$ or $k_z$ is different from $0$.

Assume $k_z\neq 0$, consider the triple $(\bar{k}-(j-1),k_y, k_z-1)$, we have that the sum of the three elements is $\bar{k}$, and also
$\bar{k}-(j-1)\geq k_y \geq k_z-1$, thus, by the inductive assumption $X^n_{(\bar{k}-(i-1),{k}_y,{k}_z-1)}\in {\mathcal{L}}$. 
We have:
\[
[X^n_{(\bar{k}-(i-1),{k}_y,{k}_z-1)}, X^n_{(0,1,0)}]= 2 X^n_{(\bar{k}-i,k_y,k_z)} -2 X^n_{(\bar{k}-i+2,k_y,k_z-2)}.
\]
Since the second element is in ${\mathcal{L}}$ by the inductive assumption, we have that $X^n_{(\bar{k}-i,k_y,k_z)} \in {\mathcal{L}}$.

If $k_z=0$, the triple is $(\bar{k}-i, i, 0)$. We compute:
\[
[X^n_{(\bar{k}-(i-1),i-1,0)}, X^n_{(0,0,1)}]=-2X^n_{(\bar{k}-i,i,0)} -2 X^n_{(\bar{k}-i+2,i-2,0)}.
\]
Again we can conclude, since the second element on the right hand side is in $\mathcal{L}$.
\epr

\end{itemize}

\section{Dynamical Lie algebra analysis for $n=2$ and $n=3$}

\subsection{Case $n=2$ and generalizations}

The symmetric vectors  
$|\phi_0\rangle$, $|\phi_1\rangle$, and $|\phi_2\rangle$, defined in (\ref{statesphi}) are an orthonormal basis of the $+1$ eigenspace of $\Pi$ in (\ref{Pimat}) while the unit (antisymmetric) vector 
\be{antisym}
|\psi_0\rangle := \frac{1}{\sqrt{2}}(-|01\rangle+|10\rangle), 
\ee
span the, $1$-dimensional, $-1$ eigenspace of $\Pi$. Writing the corresponding change of basis matrix in the computational basis, we obtain,  with $a:=\frac{1}{\sqrt{2}}$, 
\be{MatrixT}
T^\dagger:=\left[ |\psi_0 \rangle, |\phi_0 \rangle, |\phi_1 \rangle, |\phi_2 \rangle\right]:= \begin{pmatrix} 0 & 1 & 0 & 0 \cr 
-a & 0 & a & 0 \cr 
a & 0 & a & 0 \cr 
0 & 0 & 0 & 1  \end{pmatrix}.
\ee
We have, with $\Pi$ in (\ref{Pimat}) 
$T\Pi T^\dagger=diag(-1,1,1,1):={\bf 1}_{1,3}$.\footnote{In the following we shall use the notation ${\bf 1}_{j,k}$ for $diag(-1,-1,...,-1,1,1,...,1)$ with $j$ `-1's and $k$, `1's.} Therefore 
a matrix $A \in u(4)$ is in ${\cal L}^{PI}$ defined in 
(\ref{LPI}) if and only if 
$$
T \Pi T^\dagger (TAT^\dagger)T \Pi T^\dagger={\bf 1}_{1,3} (TAT^\dagger) {\bf 1}_{1,3}=
TAT^\dagger, 
$$
that is, if and only if $\tilde A:=TAT^\dagger$ commutes with ${\bf 1}_{1,3}$. This happens if and only if $\tilde A$ has a block diagonal form with two blocks of dimension $1$ and $3$. This proves that ${\cal L}^{PI}$ is $u(1) \oplus u(3)$ in this case, where the sum is a direct sum of Lie algebras (the two addenda commute) and the corresponding Lie group is the direct product of $U(1)$ and $U(3)$, a compact Lie group.

\br{TBext}
A different way to arrive at the change of coordinates $T$ in (\ref{MatrixT}), which will then be generalized to the case $n=3$ is to notice that $|\phi_0\rangle,$ $ |\phi_1\rangle$ and $|\phi_2\rangle$ span an invariant subspace for the generators $H_{zz}$, $H_x$ and $H_y$ of the Lie algebra ${\cal L}^{PI}$ and therefore for the whole Lie algebra. The same thing is true for the subspace spanned by $|\psi_0 \rangle$. Therefore, in the basis given by the matrix $T$ in (\ref{MatrixT}) the elements of ${\cal L}^{PI}$ are in the $1+3$ block diagonal form.
\er

\br{Tartaglia}
This result can be generalized in at least two ways to the case of a general number $n$ of spin. 
In particular assume we have $n$ spin and we are interested in the superalgebra of ${\cal L}^{PI}$ of matrices invariant under permutation of two of the spins, which we can assume without loss of generality to be the first two. That is,  we are interested in the superalgebra of ${\cal L}^{PI}$ 
\be{Superalg}
{\cal L}^{PI12}:=\{ A \in u(2^n) \, | \, \Pi_{1,2} A \Pi_{1,2}\}. 
\ee
Using the change of coordinates $T \otimes {\bf 1}_{2^{n-2}}$, 
we have that 
$T \otimes {\bf 1}_{2^{n-2}}\Pi_{1,2}T^\dagger \otimes {\bf 1}_{2^{n-2}}={\bf 1}_{2^{n-2}, 3\times 2^{n-2}}$ 
so that, analogously to above, modulo a change of coordinates, the matrices in ${\cal L}^{PI12}$ 
are all the matrices in $u({2^n})$ which are block diagonal with blocks of dimension $2^{n-2}$ and 
$3\times 2^{n-2}$. Therefore ${\cal L}^{PI12}=u(2^{n-2}) \oplus u(3 \times 2^{n-2})$ and the corresponding Lie group is the direct product of $U(2^{n-2})$ and $U(3 \times 2^{n-2})$, again a compact Lie group. 

Assume now $n$ even. and consider the superalgebra of ${\cal L}^{PI}$ of all the matrices $A$ such that 
\be{Leven}
{\cal L}^e:=\{A \in u(2^n) \, | \, \Pi \otimes \Pi \otimes \cdots \otimes \Pi A 
\Pi \otimes \Pi \otimes \cdots \otimes \Pi =A\}. 
\ee 
Using the change of coordinates $A \rightarrow T\otimes T \otimes \cdots \otimes T A 
T^\dagger \otimes T^\dagger \otimes \cdots \otimes T^\dagger$ we see that, in the new coordinates, the matrices in ${\cal L}^e$ are tensor products of block diagonal $4 \times 4$ matrices with one $1 \times 1$ block and one $3 \times 3$ block. So the Lie algebra ${\cal L}^e$ is spanned by all the matrices in $i (iu(1) \oplus iu(3))^{\otimes \frac{n}{2}}$. Using the fact that $iu(j) \otimes  iu(k)=iu(jk)$, we see that 
\be{Tartaglia2}
{\cal L}^e=\oplus_{j=0}^{\frac{n}{2}} { \frac{n}{2}\choose  j} i u(3^j),    
\ee
where for a positive integer $k$, we have denoted by $k u(\cdot)$ the direct sum of $k$ copies of $u(\cdot)$. 
\er 
\br{compactness}
The Lie algebra ${\cal L}^{PI}$ is the intersection of all the Lie algebras ${\cal L}^{PIjk}$ defined analogously to ${\cal L}^{PI12}$ in (\ref{Superalg}). All these Lie algebras are  conjugate (and therefore isomorphic) to   ${\cal L}^{PI12}$. Therefore the corresponding Lie group 
$e^{{\cal L}^{PI}}$ is the intersection of the compact Lie groups $e^{{\cal L}^{PIjk}}$ and therefore compact. 
\er 

Applying the change of coordinates $T$ in (\ref{MatrixT}) to 
the basis $\{ |\phi_0 \rangle, 
|\phi_1 \rangle, |\phi_2 \rangle  \}$, we obtain the basis 
\be{newbasis}
|\tilde \phi_0 \rangle:=T | \phi_0 \rangle=\begin{pmatrix} 0 \cr 1 \cr 0 \cr 0\end{pmatrix}, \qquad 
|\tilde \phi_1 \rangle:=T | \phi_1 \rangle=\begin{pmatrix} 0 \cr 0 \cr 1 \cr 0  \end{pmatrix}, \qquad 
|\tilde \phi_2\rangle:=T | \phi_2 \rangle=\begin{pmatrix} 0 \cr 0 \cr 0 \cr 1  \end{pmatrix}.  \qquad 
\ee 
Since the Lie group of block diagonal matrices, direct product of $U(1)$ and $U(3)$,  
is transitive on the manifold of linear combinations of $|\tilde \phi_0 \rangle$, $|\tilde \phi_1 \rangle$, $|\tilde \phi_2 \rangle$ 
with unit norm, (natural) pure state controllability follows. We summarize in the following Theorem: 

\bt{TheosumN2}
In the coordinates given by the matrix $T$ in (\ref{MatrixT}) the Lie algebra of permutation invariant matrices ${\cal L}^{PI}$ is made of block diagonal 
matrices with skew-Hermitian blocks of dimensions $1$ and $3$. The set of reachable evolutions is the Lie group of block diagonal matrices $diag(U_1,U_3)$ with $U_1$ ($U_3$)  unitary of dimension $1$ ($3$) and $\det(U_1)\det(U_3)=1$. System (\ref{basicHam}) is pure state controllable on the space of permutation invariant states.     
\et 

\subsection{Case $n=3$}

For $n=3$, the dynamical Lie algebra ${\cal L}$ is the 
intersection of ${\cal L}^{PI12}$ and ${\cal L}^{PI23}$ 
defined in Remark \ref{Tartaglia} inside $su(8)$. In order to find a system of coordinates where ${\cal L}^{PI}$ has a form which easily displays its Lie algebra structure we follow the idea of Remark \ref{TBext} and find orthonormal subspaces which are invariant for the generators of ${\cal L}^{PI}$ and therefore 
for all of ${\cal L}^{PI}$. One such subspace is given by 
$$
{\cal S}_\phi:=\texttt{span}\{|\phi_0 \rangle, |\phi_1 \rangle, |\phi_2 \rangle, |\phi_3 \rangle \}, 
$$   
with $|\phi_j\rangle$ defined in (\ref{statesphi}). Let 
$$|\psi_0\rangle:=-\frac{1}{\sqrt{2}}|010\rangle+\frac{1}{\sqrt{2}}|100\rangle$$ and $$|\psi_1\rangle :=-\frac{1}{\sqrt{2}}|011 \rangle +\frac{1}{\sqrt{2}} |101\rangle$$. It is a straightforward calculation to show that the subspace 
$$
{\cal S}_{\psi}:=\texttt{span}\{|\psi_0 \rangle, |\psi_1\rangle \},  
$$ 
is invariant under $H_{zz}$, $H_x$ and $H_y$ defined in (\ref{basicHam}) and (\ref{basicHamdetails}) and therefore for the whole Lie algebra ${\cal L}^{PI}$. 
Moreover consider the orthonormal vectors 
$$
|\chi_0 \rangle:=\frac{\sqrt{2}}{\sqrt{3}}|001\rangle -\frac{1}{\sqrt{6}} |010\rangle - \frac{1}{\sqrt{6}}|100\rangle, 
$$
$$
|\chi_1 \rangle := \frac{1}{\sqrt{6}}|011\rangle+\frac{1}{\sqrt{6}}|101\rangle-
\frac{\sqrt{2}}{\sqrt{3}} |110\rangle. 
$$
Again, a straightforward calculation shows that the subspace 
$$
{\cal S}_{\chi}:=\texttt{span}\{|\chi_0 \rangle, |\chi_1\rangle \},  
$$ 
is invariant under $H_{zz}$, $H_x$ and $H_y$ defined in (\ref{basicHam}) and (\ref{basicHamdetails}) and therefore for the whole Lie algebra ${\cal L}^{PI}$, moreover it is orthogonal to ${\cal S}_\phi$ and ${\cal S}_\psi$. Therefore, the matrix 
\be{Mdagger}
M^\dagger:=\left[ |\psi_0 \rangle , |\psi_1\rangle , |\chi_0\rangle , |\chi_1 \rangle , |\phi_0 \rangle , |\phi_1 \rangle , |\phi_2 \rangle , |\phi_3 \rangle\right]:=
\begin{pmatrix}      0 & 0 & 0 & 0 & 1 & 0 & 0 & 0 \cr 
0 & 0 & \frac{\sqrt{2}}{\sqrt{3}}& 0 & 0 & \frac{1}{\sqrt{3}}& 0 & 0 \cr 
-\frac{1}{\sqrt{2}} & 0 & - \frac{1}{\sqrt{6}}& 0 & 0 & \frac{1}{\sqrt{3}}& 0 & 0 \cr 
0 & - \frac{1}{\sqrt{2}}& 0 & \frac{1}{\sqrt{6}}& 0 & 0 & \frac{1}{\sqrt{3}}& 0 \cr 
\frac{1}{\sqrt{2}} & 0 & -\frac{1}{\sqrt{6}} & 0 & 0 & \frac{1}{\sqrt{3}} & 0 & 0 \cr 
0 & \frac{1}{\sqrt{2}} & 0 & \frac{1}{\sqrt{6}}& 0 & 0 & \frac{1}{\sqrt{3}}& 0 \cr 
0 & 0 & 0 & -\frac{\sqrt{2}}{\sqrt{3}}& 0 & 0 & \frac{1}{\sqrt{3}}& 0 \cr 
0 & 0 & 0&0&0&0&0& 1\end{pmatrix}, 
\ee
is such that the matrices in $M{\cal L}^{PI} M^\dagger$ have the form 
\be{MLM}
MAM^\dagger :=\begin{pmatrix} W_1 & 0 & 0 \cr 0 & W_2 & 0 \cr 0 & 0 &  W_3\end{pmatrix},  
\ee 
with $ W_1\in u(2)$, $ W_2\in u(2)$, $W_3\in u(4)$. Furthermore using the fact that $\Pi_{23}A \Pi_{23}=A$, for $A \in {\cal L}^{PI}$ and (\ref{MLM}), we obtain 
\be{MLM1}
M \Pi_{23} M^\dagger \begin{pmatrix} W_1 & 0 & 0 \cr 0 & W_2 & 0 \cr 0 & 0 & W_3\end{pmatrix} M \Pi_{23} M^\dagger=\begin{pmatrix}  W_1 & 0 & 0 \cr 0 & W_2 & 0 \cr 0 & 0 &  W_3\end{pmatrix}. 
\ee
Using $M^\dagger$  in (\ref{Mdagger}) and the (easily verifiable) relations\footnote{Notice also the relations which will not be used $\Pi_{12} |\psi_0 \rangle=-|\psi_0 \rangle$, $\Pi_{12} |\psi_1\rangle=-|\psi_1\rangle$, $\Pi_{12}|\chi_0 \rangle=|\chi_0\rangle$, $\Pi_{12} |\chi_1 \rangle=|\chi_1 \rangle$ which together with (\ref{toBadded}) show the invariance of the subspaces $\texttt{span}\{ |\psi_0\rangle, |\chi_0 \rangle \}$ and  
$\texttt{span}\{ |\psi_1\rangle, |\chi_1 \rangle \}$ under both $\Pi_{12}$ and  $\Pi_{23}$.}  
\be{toBadded}
\Pi_{23} |\psi_0\rangle = \frac{1}{2} |\psi_0 \rangle-\frac{\sqrt{3}}{2} |\chi_0\rangle, \qquad \Pi_{23} |\psi_1\rangle=\frac{1}{2} |\psi_1\rangle - \frac{\sqrt{3}}{2} |\chi_1\rangle,
\ee
$$ 
\Pi_{23} |\chi_0 \rangle =-\frac{1}{2} |\chi_0 \rangle - \frac{\sqrt{3}}{2} |\psi_0 \rangle, \qquad \Pi_{23} |\chi_1 \rangle=-\frac{1}{2} |\chi_1 \rangle -\frac{\sqrt{3}}{2} |\psi_1 \rangle, 
$$
we calculate 
\be{MPM3}
M\Pi_{23}M^\dagger:=\begin{pmatrix} \frac{1}{2} {\bf 1}_2 & -\frac{\sqrt{3}}{2} {\bf 1}_2 & 0 & 0  \cr 
 -\frac{\sqrt{3}}{2} {\bf 1}_2 & \frac{1}{2} {\bf 1}_2 & 0 & 0 \cr 
 0 & 0 & {\bf 1}_2 & 0 \cr 
 0 & 0 & 0 & {\bf 1}_2. 
 \end{pmatrix}
\ee
This, used in (\ref{MLM1}) gives $ W_1=W_2$. In conclusion in the new coordinates, matrices in ${\cal L}^{PI}$ must be of the form 
\be{Nuovaforma}
\hat B:=\begin{pmatrix} W & 0 & 0 \cr 
0 &W & 0 \cr 
0 & 0 & Q \end{pmatrix}, 
\ee
with $ W$ and $Q$ arbitrary skew-Hermitian matrices of dimensions $2$ and $4$, respectively. Since the number of degrees of freedom in (\ref{Nuovaforma}) is equal to the dimension of ${\cal L}^{PI}$ calculated in (\ref{dimensione}) (which for $n=3$ gives $20$) the Lie algebra of matrices in (\ref{Nuovaforma}) with trace equal to $0$ gives exactly ${\cal L}$. The Lie group corresponding to the dynamical Lie algebra ${\cal L}={\cal L}^{PI} \cap su(8)$, which is the space of available evolutions for the system (\ref{basicHam}) is, in the given coordinates,  the Lie group of matrices of the form 
\be{formagruppo}
\begin{pmatrix} U_2 & 0 & 0 \cr 0 & U_2 & 0 \cr 
0 & 0 & U_4, \end{pmatrix}
\ee 
with $U_2$ and $U_4$ arbitrary unitary matrices of dimensions $2$ and $4$, respectively, and $[\det(U_2)]^2\det(U_4)=1$. In the new coordinates $|\phi_0\rangle$, $|\phi_1\rangle$, 
$|\phi_2\rangle$ and $|\phi_4 \rangle$ are the elements of the standard basis $\vec e_5$, $\vec e_6$, $\vec e_7$, $\vec e_8$. From this, and the fact that $U(4)$ is transitive on the complex sphere of dimension $4$, it follows the pure state controllability of system (\ref{basicHam}) in the space of permutation invariant states. Summarizing we have the following Theorem which is the corresponding of Theorem \ref{TheosumN2} for the case $n=3$. 

\bt{TheosumN3}
In the coordinates given by the matrix $M$  in (\ref{Mdagger}), the Lie algebra of permutation invariant matrices ${\cal L}^{PI}$ is made of block diagonal 
matrices with skew-Hermitian blocks of dimensions $2$, $2$  and $4$, where the blocks of dimension $2$ are equal. The set of reachable evolutions is the Lie group of block diagonal matrices $diag(U_2,U_2,U_4)$ with $U_2$ ($U_4$)  unitary of dimension $2$ ($4$) and $\det(U_2)^2\det(U_4)=1$. System (\ref{basicHam}) is pure state controllable on the space of permutation invariant states.   
\et 
\section{Algorithms for control}\label{Algo1}
We now give algorithms for control to perform an arbitrary unitary on the space of permutation invariant states. The change of coordinates displayed in the previous 
section shows that we have a problem of control on $U(3)$ and $U(4)$ respectively. In fact the upper blocks  of the matrices in ${\cal L}$ (in the new coordinates) can be neglected since they do not affect the space of permutation invariant states. We shall assume that in (\ref{basicHam}) we can use arbitrarily large controls possibly in very short time (hard pulses). This will allows us to use methods of Cartan decompositions of Lie groups for control.

\subsection{Case $n=2$}
Consider the matrices $-iH_x$, $-iH_y$ and $-iH_{zz}$ defined in  (\ref{basicHamdetails}) for the case $n=2$. In the new coordinates defined by the matrix $T$ in (\ref{MatrixT}), $-iH_x$, $-iH_y$ and $-iH_{zz}$ transform respectively into 
$$
A_x:=T(-iH_x)T^\dagger= \begin{pmatrix} 0 & 0 & 0 & 0 \cr 
0 & 0 & -i\sqrt{2} & 0 \cr 
0 & -i\sqrt{2} & 0 & -i\sqrt{2} \cr 
0 & 0 & -i\sqrt{2} & 0 \end{pmatrix}, \quad 
A_y:=T(-iH_y)T^\dagger= \begin{pmatrix} 0 & 0 & 0 & 0 \cr 
0 & 0 & \sqrt{2} & 0 \cr 
0 & -\sqrt{2} & 0 & \sqrt{2} \cr 
0 & 0 & -\sqrt{2} & 0 \end{pmatrix}, \quad 
$$
$$
A_{zz}:=T(-iH_{zz})T^\dagger=\begin{pmatrix}i & 0 & 0 & 0 \cr 0 & -i & 0 & 0 \cr 
0 & 0 & i & 0 \cr 
0 & 0 & 0 & -i  \end{pmatrix}. 
$$ 
One extra change of coordinates $A \rightarrow \hat T A \hat T^\dagger$ with 
\be{hatTmatrix}
\hat T:=\begin{pmatrix} 1 & 0 & 0 & 0 \cr 
0& \frac{-i}{\sqrt{2}} & 0 & \frac{-i}{\sqrt{2}} \cr 
0 & 0 & 1 & 0 \cr 
0 & \frac{1}{\sqrt{2}} & 0 & \frac{-1}{\sqrt{2}}  \end{pmatrix},  
\ee 
gives 
$$
\hat A_x:=\hat T(A_x)\hat T^\dagger= \begin{pmatrix} 0 & 0 &0 & 0 \cr 
0 & 0 &-2 & 0 \cr 
0 & 2 & 0 & 0 \cr 
0 & 0 & 0 & 0 
\end{pmatrix}, \quad 
\hat A_y:=\hat T(A_y)\hat T^\dagger= \begin{pmatrix} 0 & 0 & 0 & 0 \cr 
0 & 0 & 0 & 0 \cr 
0 & 0 & 0 & -2 \cr 
0 & 0 & 2 & 0
\end{pmatrix}, \quad 
$$
$$
\hat A_{zz}:=\hat T(A_{zz}) \hat T^\dagger=
\begin{pmatrix} i & 0 & 0 & 0 \cr 
0 & -i & 0 & 0 \cr 
0 & 0 & i & 0 \cr 
0 & 0 & 0 & -i  \end{pmatrix}. 
$$ 
In these coordinates the system (\ref{basicHam}) becomes the right invariant system on a Lie group 
\be{riscon}
\dot X=\hat A_z X+ \hat A_xXu_x+ \hat A_yXu_y.
\ee
Neglecting the upper $1\times 1$ block of the matrix $X$ (which does not affect the permutation invariant states), and neglecting matrices which are multiples of the identity which only add a phase factor to the solution,   this system has a $P-K$ structure, i.e., there exists a Cartan decomposition of $su(3)={\cal K} \oplus {\cal P}$, with 
$$
[{\cal K}, {\cal K}]\subseteq {\cal K}, \qquad [{\cal K}, {\cal P}]\subseteq {\cal P}, \qquad
[{\cal P}, {\cal P}]\subseteq {\cal K}
$$
such that the matrices multiplying the control (in this case $\hat A_x$ and $\hat A_y$)  
 generate the Lie subalgebra ${\cal K}$ and the drift matrix (in this case $\hat A_z$) belongs to ${\cal P}$. In our case, the decomposition 
 is obtained with ${\cal K}=so(3)$ while ${\cal P}$ is the space of purely imaginary matrices. The method of control in this case is as follows:

 First  write the desired final condition $X_f \in SU(3)$ according to the Cartan decomposition as $X_f:=K_1 A K_2$ where $K_1$ and $K_2$ which are in the Lie group corresponding to ${\cal K}$ (in this case $SO(3)$). The matrix $A$ is an element 
 of the Lie group associated to a Cartan subalgebra (i.e., a maximal Abelian subalgebra contained in ${\cal P}$). Then the problem is to obtain  evolutions which implement $K_2$, $A$ and $K_1$ in that order. $K_1$ and $K_2$ are obtained with hard pulses, high amplitude short time controls, which essentially allow us  to neglect the drift term.  
 The element $A$  is implemented by alternating hard pulses with free evolutions (setting the controls equal to zero).

 Details of the  approach to control based on decompositions of Lie groups are described in  \cite{Mikobook} which also gives computational methods to find the factors $K_{1,2}$ and $A$ in the decomposition. The paper \cite{Khan1} 
 shows that this method of control is in fact time optimal.

\subsection{Case $n=3$} 

Analogously to the case $n=2$, we first transform $-iH_x$, 
$-iH_y$ and $-iH_{zz}$ in new coordinates using the transformation $M$  in (\ref{Mdagger}). A direct calculation (using Appendix B) shows: 
\be{MHxMd}
M(-iH_x)M^\dagger:=-i\begin{pmatrix} 0 & 1 & 0 & 0 & 0 &0 & 0 & 0 \cr 
1 & 0 & 0& 0& 0& 0& 0 & 0 \cr 
0 & 0 & 0 & 1 & 0 & 0  & 0 & 0 \cr
0 & 0 & 1 & 0 & 0 & 0  & 0 & 0 \cr
0 & 0 & 0 & 0 & 0 & \sqrt{3}  & 0 & 0 \cr
0 & 0 & 0&0&\sqrt{3} & 0  & 2 & 0 \cr
0 & 0 & 0 & 0 & 0 & 2 & 0 & \sqrt{3} \cr 
0 & 0 & 0 & 0 & 0 & 0 & \sqrt{3} & 0 \end{pmatrix} 
\ee

\be{MHyMd}
M(-iH_y)M^\dagger:=\begin{pmatrix} 0 & 1 & 0 & 0 & 0 &0 & 0 & 0 \cr 
-1 & 0 & 0& 0& 0& 0& 0 & 0 \cr 
0 & 0 & 0 & 1 & 0 & 0  & 0 & 0 \cr
0 & 0 & -1 & 0 & 0 & 0  & 0 & 0 \cr
0 & 0 & 0 & 0 & 0 & \sqrt{3}  & 0 & 0 \cr
0 & 0 & 0&0&-\sqrt{3} & 0  & 2 & 0 \cr
0 & 0 & 0 & 0 & 0 & -2 & 0 & \sqrt{3} \cr 
0 & 0 & 0 & 0 & 0 & 0 & -\sqrt{3} & 0 \end{pmatrix} 
\ee

\be{MHzzMd}
M(-iH_{zz})M^\dagger:=-i\begin{pmatrix} -1 & 0 & 0 & 0 & 0 &0 & 0 & 0 \cr 
0 & -1 & 0& 0& 0& 0& 0 & 0 \cr 
0 & 0 & -1 & 0 & 0 & 0  & 0 & 0 \cr
0 & 0 & 0 & -1 & 0 & 0  & 0 & 0 \cr
0 & 0 & 0 & 0 & 3& 0  & 0 & 0 \cr
0 & 0 & 0&0&0 & -1 & 0 & 0 \cr
0 & 0 & 0 & 0 & 0 & 0 & -1 &0 \cr 
0 & 0 & 0 & 0 & 0 & 0 &0 & 3 \end{pmatrix} 
\ee

From now on we shall only focus on the last four rows and columns which represent the evolution of the dynamics on the subspace of permutation invariant states spanned by $|\phi_0 \rangle$, $|\phi_1 \rangle$, $|\phi_2 \rangle$, and $|\phi_3 \rangle$. The problem of control is solved if we are able to factorize the desired final evolution $X_f \in U(4)$ in exponentials of matrices proportional to (cf. (\ref{MHxMd})), (\ref{MHyMd}), (\ref{MHzzMd})) 
\be{BxByBz}
B_x:=\begin{pmatrix} 0 & \sqrt{3} i & 0 & 0 \cr 
\sqrt{3} i & 0 & 2i & 0 \cr 
0 & 2i & 0 & \sqrt{3}i \cr 
0 & 0 & \sqrt{3} i & 0  \end{pmatrix}, \qquad B_y:=\begin{pmatrix} 0 & \sqrt{3} & 0 & 0 \cr 
- \sqrt{3} & 0 & 2 & 0 \cr 
0 & -2 & 0 & \sqrt{3} \cr 
0 & 0 & -\sqrt{3} & 0  \end{pmatrix}, \qquad \tilde B_{zz}:=
\begin{pmatrix}-3i & 0 & 0 & 0 \cr 0 & i & 0 & 0 \cr 0 & 0 & i & 0 \cr 0 & 0 & 0 & -3i \end{pmatrix}, 
\ee
that is as exponentials of the form $e^{B_xt}$, $e^{B_y t}$ and $e^{\tilde B_{zz}t}$ for real $t$. The exponentials of the form $e^{B_xt}$ and  $e^{B_y t}$ can be obtained using hard pulses in the Hamiltonian  (\ref{basicHam}), the elements 
$e^{\tilde B_{zz}t}$ are obtained by setting the controls equal to zero and allowing the system free evolution. Notice that the orbit $\{  e^{\tilde B_{zz}t}\, | \, t\in \RR\}$ is periodic and so we can obtain all the values in it even if we restrict ourselves to positive values of the time $t$ as it is required in physical applications.   Furthermore, by neglecting an overall phase factor which does not have a physical meaning,and rescaling the time $t$,  we can consider instead of the matrix $\tilde B_{zz}$ in (\ref{BxByBz}), the matrix 
\be{Bz}
B_{zz}:=-\frac{1}{2} \left(\tilde B_{zz}+i {\bf 1}\right)=\begin{pmatrix} 
 i & 0 & 0 & 0 \cr 0 & -i & 0 & 0 
\cr 0 & 0 & -i & 0 \cr 0 & 0 & 0 & i\end{pmatrix},  
\ee
and restrict to matrices $X_f \in SU(4)$. 

We shall again use an appropriate  Cartan decomposition along with the method for generating `new directions' described in \cite{Mikounderrated}. In particular, 
we use the {\bf AIII} KAK Cartan decomposition \cite{Helgason} of $SU(4)$ in that every element $X_f \in SU(4)$ can be factorized as 
\be{factr1}
X_f:=K_1 A K_2, 
\ee   
where $K_1$ and $K_2$ are matrices  with elements at the intersection of rows and columns 1-4 and 2-3 occupied by $2 \times 2$ unitary matrices $U_1$ and $U_2$ (by permuting row and column indexes these would be block diagonal matrices with $2\times 2$ blocks) and $\det(U_1)\det(U_2)=1$. The matrix $A$ is the product of two commuting matrices belonging to the associated Cartan subalgebra which we can take equal to 
$\texttt{span} \{ A_4, C_4\}$ with 
\be{A4C4}
A_3:=\frac{1}{2}\begin{pmatrix} 0 & 0 & 1 & 0 \cr 
0 & 0 & 0 & -1 \cr 
-1 & 0 & 0 & 0 \cr 
0 & 1 & 0 & 0 \end{pmatrix}, \qquad C_3:=\begin{pmatrix} 0 & 0 & 1 & 0 \cr 
0 & 0 & 0 & 1 \cr 
-1 & 0 & 0 & 0 \cr 
0 & -1 & 0 & 0 \end{pmatrix}, 
\ee    
which are commuting. Methods to compute the factors in (\ref{factr1}) are described in \cite{Mikobook}.  Our task is therefore to show how to express matrices of the form $K_1$ $K_2$ and $A$ in (\ref{factr1}) as products of exponentials of the matrices $B_x$, $B_y$ and $B_{zz}$.  In order to do that we shall consider two Lie subalgebras of $su(4)$ which are isomorphic to each other. In particular consider the Lie algebra ${\cal A}$ given by 
$$
{\cal A}:=\texttt{span}\{A_1, A_2, A_3, E\},  
$$
with $A_3$ given in (\ref{A4C4}) and  
\be{LIEA}
A_1:=\frac{1}{2}\begin{pmatrix} 0 & 1 & 0 & 0 \cr 
-1 & 0 & 0 & 0 \cr 
0 & 0 & 0 & 1 \cr 
0 & 0 & -1 & 0 \end{pmatrix}, \quad
 A_2:=\frac{1}{2}\begin{pmatrix}0 & 0 & 0 & 1 \cr 
0 & 0 & 1 & 0 \cr 
0 & -1 & 0 & 0 \cr 
-1 & 0 & 0 & 0 \end{pmatrix},  E:=\frac{1}{2}\begin{pmatrix} 0 & 0 & 0 & 1 \cr 
0 & 0 & -1 & 0 \cr 
0 & 1 & 0 & 0 \cr 
- 1& 0 & 0 & 0  \end{pmatrix}. 
\ee
We have the commutation relations 
\be{commurel45}
[A_1, A_2]=A_3, \qquad [A_2, A_3]=A_1, \quad [A_3, A_1]=A_2, \quad [{\cal A}, E]=0, 
\ee
which show that ${\cal A}$ is the direct sum of a Lie subalgebra isomorphic to $su(2)$ with its centralizer spanned by $E$. 

Consider now the Lie algebra ${\cal B}$, with ${\cal B}:=\texttt{span}\{ B_1, B_2, B_3, F\}$ where $B_1:=A_1$ in (\ref{LIEA}) and 
\be{LIEB}
B_2:=\frac{1}{2} \begin{pmatrix}0 & 0 & 0 & i \cr 
0 & 0 & i & 0 \cr 
0 & i & 0 & 0 \cr
i & 0 & 0 & 0 \end{pmatrix}, \quad B_3:=\frac{1}{2} \begin{pmatrix} 
0 & 0 & i & 0 \cr 
0 & 0 & 0 & -i \cr 
i & 0 &0 & 0 \cr 
0 & -i & 0 & 0 
\end{pmatrix}, \quad F:=\frac{1}{2}\begin{pmatrix}0 & 0 & 0 & i \cr 
0 & 0 & -i & 0 \cr 
0 & -i & 0 & 0 \cr 
i& 0 & 0 & 0   \end{pmatrix}, 
\ee
and we have the commutation relations 
\be{commurel46}
[B_1, B_2]=B_3, \qquad [B_2, B_3]=B_1, \quad [B_3, B_1]=B_2, \quad [{\cal B}, F]=0, 
\ee
which, compared to (\ref{commurel45}) show that ${\cal B}$ 
is isomorphic to ${\cal A}$.

We shall assume first that we are able to obtain all matrices in the connected Lie groups corresponding to ${\cal A}$ and ${\cal B}$, i.e., $e^{\cal A}$ and $e^{\cal B}$ and show that with these we can construct the decomposition (\ref{factr1}). Then we will show how to obtain any element on these Lie groups. 

The matrices of the form $e^{(A_2-E)t} \in e^{\cal A}$ 
($e^{(A_2+E)t} \in e^{\cal A}$)  are equal to the identity except for the rows and columns $2$ and $3$ ($1$ and $4$) which contain an arbitrary (depending on $t$) $Y-$rotation. Analogously, the matrices of the form $e^{(B_2-F)t} \in e^{\cal B}$ ($e^{(B_2-F)t} \in e^{\cal B}$)  are equal to the identity except for the rows and columns $2$ and $3$ ($1$ and $4$) which contain an arbitrary (depending on $t$) $X-$rotation. Using these matrices and  Euler decompositions we obtain matrices such that the elements corresponding to indexes $1$ and $4$ give an arbitrary matrix in $SU(2)$, and the elements corresponding to indexes $2$ and $3$ give an arbitrary matrix in $SU(2)$. Multiplying the overall $4 \times 4$ matrix   by a matrix of the form $e^{B_{zz}t}$ we obtain matrices of the form $K_1$ and $K_2$ in (\ref{factr1}). 
The element $A$ in (\ref{factr1}) is obtained as the product of an element of the form $e^{A_3 t}$ and an element of the form $e^{C_3t}$ (cf. (\ref{A4C4}) which commute. $e^{A_3 t} in e^{\cal A}$, while, it is immediate to verify that 
$e^{-B_{zz} \frac{\pi}{4}} B_3 e^{B_{zz} \frac{\pi}{4}}=C_3$, so that 
$$
e^{C_3t}=e^{-B_{zz} \frac{\pi}{4}} e^{B_3 t}  e^{B_{zz} \frac{\pi}{4}}. 
$$

Finally, we show how to obtain arbitrary elements in $e^{\cal A}$ and $e^{\cal B}$. We start with  $e^{\cal A}$ ($e^{\cal B}$ is similar). We are allowed to take exponentials $e^{B_yt}$ for $B_y$ in (\ref{BxByBz}) but also exponentials of 
\be{hatby}
\hat B_y:=e^{B_{zz} \frac{\pi}{2}} B_y e^{-B_{zz} \frac{\pi}{2}}=\begin{pmatrix}0 & -\sqrt{3} & 0 & 0 \cr 
 \sqrt{3} & 0 & 2 & 0 \cr 
0 & -2 & 0 & -\sqrt{3} \cr 
0 & 0 & \sqrt{3} & 0  \end{pmatrix}.  
\ee
In a basis given by eigenvectors of $B_y$, we write $B_y$ as 
\be{TByT}
T B_yT^\dagger:=\begin{pmatrix} 2i & 0 & 0 & 0 \cr 
0 & -2i & 0 & 0 \cr 
0 & 0 & 2i & 0 \cr 
0 & 0 & 0 & -2i  \end{pmatrix}+\begin{pmatrix}-i & 0 & 0 & 0 \cr 
0 & -i & 0 & 0 \cr 
0 & 0 & i & 0 \cr 
0 & 0 & 0 & i \end{pmatrix}. 
\ee
In the same coordinates, $\hat B_y$ becomes 
\be{ThatByT}
T \hat B_yT^\dagger:=\begin{pmatrix} -i & \sqrt{3}  & 0 & 0 \cr 
-\sqrt{3} & i & 0 & 0 \cr 
0 & 0 & -i & \sqrt{3} \cr 
0 & 0 & -\sqrt{3} & i  \end{pmatrix}+\begin{pmatrix}-i & 0 & 0 & 0 \cr 
0 & -i & 0 & 0 \cr 
0 & 0 & i & 0 \cr 
0 & 0 & 0 & i \end{pmatrix}. 
\ee
The second matrix in (\ref{TByT}) and (\ref{ThatByT}) 
commutes with both matrices and therefore spans the centralizer of the Lie algebra generated by the two matrices which is conjugate to ${\cal A}$. In these coordinates, it is also clear that such a Lie algebra is isomorphic to the direct sum of 
$su(2)$ and a one dimensional centralizer. The $2 \times 2$ blocks of the first matrices on the right hand sides in (\ref{TByT}) and (\ref{ThatByT}) are equal to each other. The problem of factorization  is therefore a problem on $SU(2)$ with an additional phase factor which we would like to fix arbitrarily. This problem can be solved by first neglecting the second term in (\ref{TByT}) and (\ref{ThatByT}) and considering the problem of factorization of elements in $SU(2)$ with matrices $e^{Z_1 t}$ and $e^{Z_2t}$, with 
\be{Z1Z2}
Z_1:=\begin{pmatrix}  2i & 0 \cr 0 & -2i\end{pmatrix}, \qquad Z_2:=\begin{pmatrix} - i & \sqrt{3}  \cr - \sqrt{3} & i \end{pmatrix}.  
\ee
This problem can be solved (with minimum number of switches) with the method described in \cite{MikoEulgen} (see also \cite{Lowenthal}). The extra phase factor can be `canceled'  by introducing and extra identity matrix   which again can be obtained in arbitrary time. In particular assume we want to obtain the matrix with $e^{-i mu} X_f$ in the upper block and $e^{i mu} X_f$ in the lower block, for $X_f \in SU(2)$. Let $\prod_{j=1} e^{Z_1 t_{j1}} e^{Z_2 t_{j2}}$ the sequence which gives $X_f$ according to the method of \cite{MikoEulgen}.  Moreover let  $\prod_{j=1} e^{Z_1 a_{j1}} e^{Z_2 a_{j2}}$ the sequence which gives $\begin{pmatrix} 0 & 1 \cr -1 & 0 \end{pmatrix}$ and $\prod_{j=1}e^{Z_1 b_{j1}} e^{Z_2 b_{j2}}$ the sequence 
which gives $\begin{pmatrix} 0 & -1 \cr 1 & 0 \end{pmatrix}$. We use 
\be{prodotto}
\prod_{j=1} e^{T  B_yT^\dagger t_{j1}} e^{T \hat B_yT^\dagger t_{j2}} 
\times \prod_{j=1} e^{T  B_yT^\dagger a_{j1}} e^{T \hat B_yT^\dagger a_{j2}}
\times e^{TB_yT^\dagger \alpha}  \prod_{j=1} e^{T  B_yT^\dagger b_{j1}} e^{T \hat B_yT^\dagger b_{j2}} \times e^{TB_yT^\dagger \alpha}.  
\ee
Set $T:=\sum_{j} (t_{j1}+ t_{j2}) + \sum_{j} (a_{j1}+ a_{j2})+ \sum_{j} (b_{j1}+ b_{j2})$
In the upper (lower) block, this give $X_f e^{-iT} e^{-2i\alpha}$ 
($X_f e^{iT} e^{2i\alpha}$) and choosing $2\alpha+T=\mu$ we obtain the desired final condition.

The treatment of $e^{\cal B}$ is perfectly analogous starting with $B_x$ in (\ref{BxByBz}) rather than $B_y$ and obtaining the extra `direction' (cf. (\ref{hatby}) 
\be{hatbx}
\hat B_x:=e^{B_{zz} \frac{\pi}{2}} B_x e^{-B_{zz} \frac{\pi}{2}}=
\begin{pmatrix}0 & -\sqrt{3} i & 0 & 0 \cr 
 -\sqrt{3}i & 0 & 2i & 0 \cr 
0 & 2i & 0 & -\sqrt{3} i \cr 
0 & 0 & -\sqrt{3}i & 0  \end{pmatrix},  
\ee  
we express the matrices $B_x$ and $\hat B_x$ in the coordinates given by the eigenvectors of $B_x$, that is for an appropriate matrix $U$ we have (cf. (\ref{TB_yT}) and (\ref{ThatByT}))
$$
UB_x U^\dagger =\begin{pmatrix} 2 i & 0 & 0 & 0 \cr 0 & -2i & 0 & 0 \cr 
0 & 0 & 2i & 0 \cr 
0 & 0 & 0 & -2i\end{pmatrix}+ \begin{pmatrix} -i & 0 & 0 & 0 \cr 0 & -i & 0 & 0 \cr 
0 & 0 & i & 0 \cr 
0 & 0 & 0 & i\end{pmatrix},  
$$
$$
U\hat B_x U^\dagger =\begin{pmatrix} -i & \sqrt{3}  & 0 & 0 \cr \sqrt{3}i & i & 0 & 0 \cr 
0 & 0 & -i & sqrt{3}i  \cr 
0 & 0 & \sqrt{3}i  & i\end{pmatrix}+ \begin{pmatrix} -i & 0 & 0 & 0 \cr 0 & -i & 0 & 0 \cr 
0 & 0 & i & 0 \cr 
0 & 0 & 0 & i\end{pmatrix}.   
$$
Then the treatment follows, as for the case of $e^{\cal A}$ from the results on factorizations of $SU(2)$ in \cite{MikoEulgen}. 

\vs

\vs

\noindent {\bf Acknowledgement} D. D'Alessandro research is supported by NSF under Grant 17890998

\section*{Appendix A: Proof of formula (\ref{forApp})}

To prove the first one, let first observe that:
\[
[{\bf{1}}\otimes\cdots{\bf{1}}\otimes\underbrace{\sigma_y}_{j^{th} }\otimes\cdots\otimes\underbrace{\sigma_z}_{l^{th} }\otimes\cdots {\bf{1}} \otimes \cdots{\bf{1}},\, {\bf{1}}\otimes\cdots{\bf{1}}\otimes\underbrace{\sigma_x}_{i^{th} }\otimes{\bf{1}}\otimes\cdots{\bf{1}}]=
\]
\be{l31}
=\left\{\begin{array}{cc}
0   & \text{ if $i\neq j$ and $i\neq l$}\\
2i \ {\bf{1}}\otimes\cdots{\bf{1}}\otimes\underbrace{\sigma_z}_{j^{th} }\otimes\cdots\otimes\underbrace{\sigma_z}_{l^{th} }\otimes\cdots {\bf{1}} \otimes \cdots{\bf{1}}
&  \text{ if $i=j$} \\
-2i \ {\bf{1}}\otimes\cdots{\bf{1}}\otimes\underbrace{\sigma_y}_{j^{th} }\otimes\cdots\otimes\underbrace{\sigma_y}_{l^{th} }\otimes\cdots {\bf{1}} \otimes \cdots{\bf{1}}
&  \text{ if $i=l$} 
\end{array}\right.
\ee
Using the previous equation we have:
\[
[ X^n_{(0,1,1)},X^n_{(1,0,0)} ] =-\sum_{\begin{array}{l} i,j,l=1\\ j\neq l \end{array}}^n 
[{\bf{1}}\otimes\cdots{\bf{1}}\otimes\underbrace{\sigma_y}_{j^{th} }\otimes\cdots\otimes\underbrace{\sigma_z}_{l^{th} }\otimes\cdots {\bf{1}} \otimes \cdots{\bf{1}},\, {\bf{1}}\otimes\cdots{\bf{1}}\otimes\underbrace{\sigma_x}_{i^{th} }\otimes{\bf{1}}\otimes\cdots{\bf{1}}]
\]
\[
= -
\sum_{\begin{array}{l} j,l=1\\ j\neq l \end{array}}^n 
2i \left( {\bf{1}}\otimes\cdots{\bf{1}}\otimes\underbrace{\sigma_z}_{j^{th} }\otimes\cdots\otimes\underbrace{\sigma_z}_{l^{th} }\otimes\cdots {\bf{1}} \otimes \cdots{\bf{1}}
-  {\bf{1}}\otimes\cdots{\bf{1}}\otimes\underbrace{\sigma_y}_{j^{th} }\otimes\cdots\otimes\underbrace{\sigma_y}_{l^{th} }\otimes\cdots {\bf{1}} \otimes \cdots{\bf{1}}\right)
\]
\[  -4\, X^n_{(0,0,2)} +4 \, X^n_{(0,2,0)} .
\]
The second equality can be proved in a similar  way.

\section*{Appendix B: Calculation of the action of $H_x$, $H_y$ and $H_{zz}$ on the basis $\{ |\psi_0 \rangle, |\psi_1 \rangle, |\chi_0 \rangle, |\chi_1 \rangle, 
|\phi_0 \rangle, |\phi_1 \rangle, |\phi_2 \rangle, |\phi_3 \rangle\}$}

\begin{eqnarray}
H_x  |\psi_0\rangle=|\psi_1\rangle \nonumber \\
H_x  |\psi_1\rangle=|\psi_0\rangle \nonumber\\
H_x  |\chi_0\rangle=|\chi_1\rangle \nonumber \\
H_x  |\chi_1\rangle=|\chi_0\rangle \nonumber \\
H_x  |\phi_0\rangle=\sqrt{3}|\phi_1\rangle \nonumber \\
H_x  |\phi_1\rangle=\sqrt{3} |\phi_0\rangle + 2 |\phi_2\rangle \nonumber \\
H_x  |\phi_2\rangle=\sqrt{3} |\phi_3\rangle + 2 |\phi_1\rangle \nonumber \\
H_x  |\phi_3\rangle=\sqrt{3} |\phi_2\rangle  \nonumber \\
\end{eqnarray}

\begin{eqnarray}
H_y  |\psi_0\rangle=-i|\psi_1\rangle \nonumber \\
H_y  |\psi_1\rangle=i|\psi_0\rangle \nonumber\\
H_y  |\chi_0\rangle=-i|\chi_1\rangle \nonumber \\
H_y  |\chi_1\rangle=i|\chi_0\rangle \nonumber \\
H_y  |\phi_0\rangle=-i\sqrt{3}|\phi_1\rangle \nonumber \\
H_y  |\phi_1\rangle=i\sqrt{3} |\phi_0\rangle - 2 i|\phi_2\rangle \nonumber \\
H_y  |\phi_2\rangle=-i\sqrt{3} |\phi_3\rangle + 2 i|\phi_1\rangle \nonumber \\
H_y  |\phi_3\rangle=i\sqrt{3} |\phi_2\rangle  \nonumber \\
\end{eqnarray}

\begin{eqnarray}
H_{zz}  |\psi_0\rangle=-|\psi_0\rangle \nonumber \\
H_{zz}   |\psi_1\rangle=-|\psi_1\rangle \nonumber\\
H_{zz}  |\chi_0\rangle=-|\chi_0\rangle \nonumber \\
H_{zz}   |\chi_1\rangle=-|\chi_1\rangle \nonumber \\
H_{zz}  |\phi_0\rangle= 3|\phi_0\rangle \nonumber \\
H_{zz}  |\phi_1\rangle=- |\phi_1\rangle \nonumber \\
H_{zz}  |\phi_2\rangle=- |\phi_2\rangle \nonumber \\
H_{zz}  |\phi_3\rangle= 3|\phi_3\rangle  \nonumber \\
\end{eqnarray}

\end{document}